\documentclass[10pt]{article}
\usepackage[T1]{fontenc}
\usepackage[utf8]{inputenc}
\usepackage{graphicx}
\usepackage{listings}
\usepackage{bm}
\usepackage{latexsym}
\usepackage{amsmath}
\usepackage{amssymb}
\usepackage{booktabs}
\usepackage{array}
\usepackage[UKenglish]{babel}
\usepackage[a4paper,top=3.5cm,bottom=3.5cm,left=3.5cm,right=3.5cm]{geometry}
\usepackage{lineno}
\usepackage{tabularx}
\usepackage{longtable}
\RequirePackage[colorlinks=true
,urlcolor=blue
,anchorcolor=blue
,citecolor=blue
,filecolor=blue
,linkcolor=blue
,menucolor=blue
,pagecolor=blue
,linktocpage=true
,pdfproducer=medialab
]{hyperref}

\usepackage{makeidx}
\newcommand{\be}{\begin{equation}}  
\newcommand{\ee}{\end{equation}}  
\newcommand{\bea}{\begin{eqnarray}}  
\newcommand{\eea}{\end{eqnarray}}  
\makeindex

\makeatletter
\g@addto@macro\bfseries{\boldmath}
\makeatother

\newcommand{\MYhref}[3][blue]{\href{#2}{\color{#1}{#3}}}
\usepackage[skip=-5pt]{caption}

\begin{document}


\begin{center}
{\LARGE \bf PDFs determination from the LHeC}

\par\vspace*{2.5mm}\par

{

\bigskip

\large \bf Francesco Giuli\footnote{E-Mail: \MYhref{francesco.giuli@cern.ch}{francesco.giuli@cern.ch}} (on behalf of the LHeC/FCC-eh study groups)}

\vspace*{2.5mm}

{CERN, EP Department, CH-1211 Geneva 23, Switzerland}

\vspace*{2.5mm}

{\it Presented at XXIX Cracow Epiphany Conference, Cracow, Poland, January 16-19 2023}

\vspace*{2.5mm}

\end{center}

\begin{abstract}
Deep Inelastic Scattering would be brought into the unexplored TeV regime by the the proposed Large Hadron Electron Collider at CERN. Its rich physics program, includes both precision Standard Model measurements to complement LHC physics as well as studies of QCD in the high energy limit. The present proceeding reports on studies included in the updated LHeC Conceptual Design Report. We study the impact of LHeC simulated data on Parton Distribution Functions uncertainties. We also assess the LHeC potential to allow the determination of the strong coupling constant $\alpha_{S}$, at per-mille level as well as to disentangle between various scenarios of small-$x$ QCD.
\end{abstract}
 
\section{The Large Hadron Electron Collider}
The Conceptual Design Report (CDR) of the Large Hadron Electron Collider (LHeC) was published in 2012~\cite{LHeCStudyGroup:2012zhm}.  The advantage of an $ep$ collider is that it offers the opportunity to observe phenomena which would be observed in a $pp$ collider with a cleaner decay environment.  The LHeC would provide a 60 GeV energy electron beam which would be collided with a hadron beams already from the LHC. In this way, there would be an increase of two orders of magnitudes over the integrated HERA luminosity,  and the kinematic range would be increased by a factor $\sim$ 25 for $Q^{2}$ and $1/x$.\\
The LHeC experiment could be realized in the 2030s during the High-Luminosity LHC (HL-LHC) and it will provide an unprecedented resolution of the partonic constituents in hadronic matter up to $x\sim$ 0.9 and down to Bjorken $x\sim$ 10$^{-6}$.

\section{Simulated samples}
In order to estimate the impact of LHeC data on the uncertainties of Parton Distribution Functions (PDFs), several sets of LHeC inclusive Neutral-Current and Charged-Current Deep Inelastic Scattering (NC and CC DIS) data have been simulated,  with a full set of systematic uncertainties. The largest source of uncertainties are due to the hadronic energy scale, global efficiency uncertainty and photo-production background. An improvement by at most a factor of two with respect to H1's achievements is foreseen.\\
The bulk of the data is assumed to be taken with electrons, characterized by large negative helicity $P_e$. It is assumed there will be an initial phase during which the LHeC may collect 50 fb$^{-1}$ of data, (first three years) while the total luminosity is assumed to be close to 1 ab$^{-1}$, a very high value when compared with HERA.

\section{PDFs determination}
The expected sensitivity of LHeC data on PDFs determinations was presented in Ref.~\cite{LHeC:2020van}.  Fits in NNLO QCD have been performed to the above-described simulated data. The HERAPDF procedure~\cite{H1:2015ubc} has been followed closely, and the following parametric functions have been adopted. The quark distributions at the initial scale $Q_{0}^{2}$ = 1.9~GeV$^{2}$ are parametrised by:
\begin{equation}
xq_{i}(x) = A_{i}x^{B_{i}}(1-x)^{C_{i}}P_{i}(x),
\end{equation}
where $P_{i}(x)=(1+D_{i}x+E_{i}x^{2})$ and the flavour of the quark distribution is specified by the index $i$. The $A_{u_{V}}$ and $A_{d_{V}}$ parameters are fixed using the quark counting rule,  while the normalisation and the slope of the $\bar{u}$ and $\bar{d}$ distributions are set equal for $x\rightarrow 0$. The strange quark PDF, $x\bar{s}$, is set to be a fraction of the $x\bar{d}$ distributions, namely $x\bar{s}=r_{s}x\bar{d}$ with $r_{s}=$ 0.67.\\
The gluon PDF, $xg$, uses a different parametrisation:
\begin{equation}
xg(x) = A_{g}x^{B_{g}}(1-x)^{C_{g}}-A^{'}_{g}x^{B^{'}_{g}}(1-x)^{C^{'}_{g}}.
\end{equation}
The momentum sum rule fixes $A_{g}$, the normalisation parameter,  while $C^{'}_{g}$ is set to 25 to suppress negative contributions at high $x$.\\
The parametrised PDFs at the starting scale $Q_{0}^{2}$ are the valence distributions, $xu_{V}$ and $xd_{V}$, the gluon distribution, $xg$, and the sea distributions $x\bar{U}$ and $x\bar{D}$, being $x\bar{U}=x\bar{u}$ and $x\bar{D}=x\bar{d}+x\bar{s}$. The nominal fits have 14 free parameters,  which are similar to HERAPDF2.0, albeit more flexible due to the stronger constraints from the LHeC.
\subsubsection{Valence distributions}
\begin{figure}[t!]
\begin{center}
\includegraphics[width=0.95\textwidth]{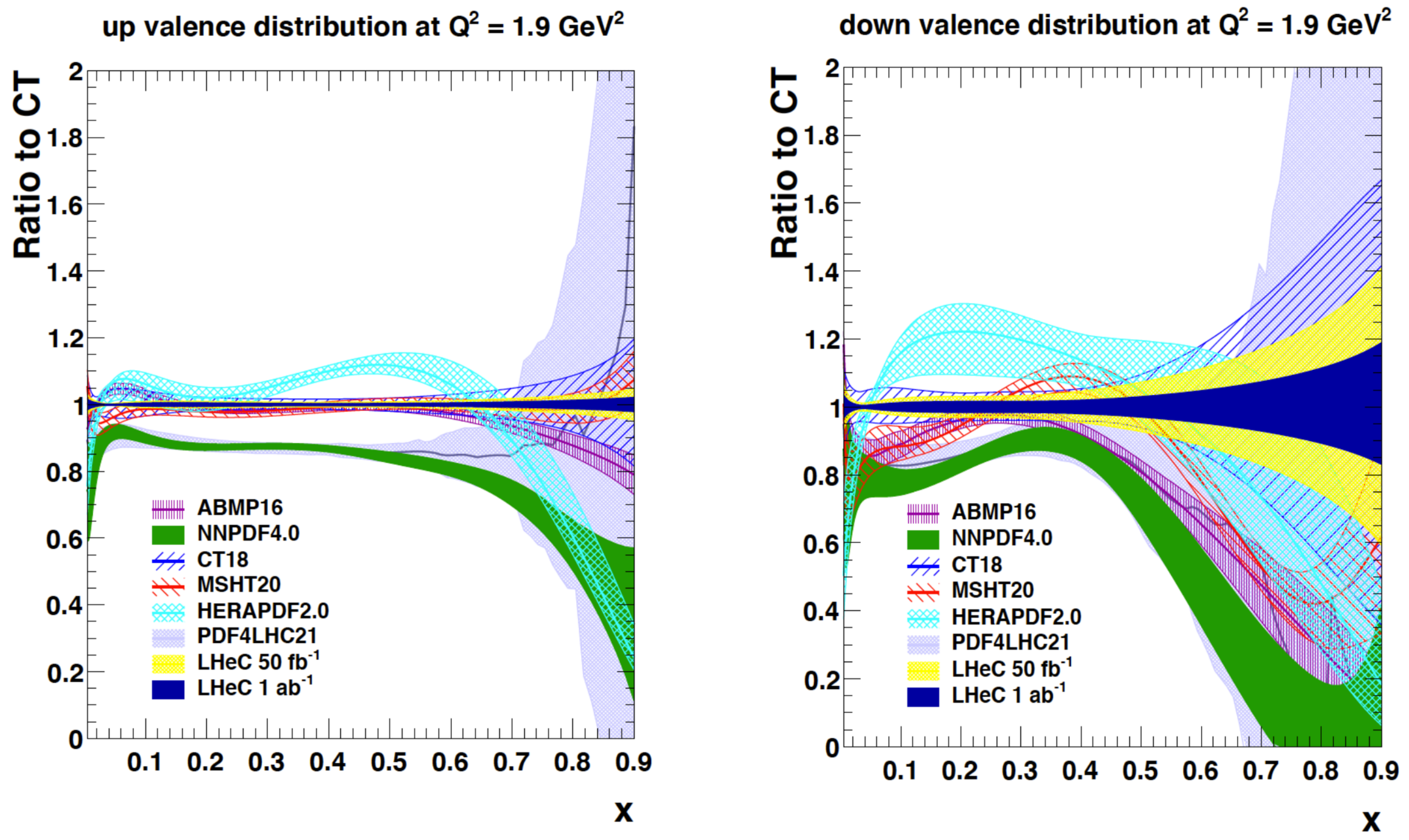}
\end{center}
\caption{Valence quark distributions at $Q^{2}$ = 1.9~GeV$^{2}$ as a function of $x$, displayed as a ratio to the CT18~\cite{Hou:2019efy} PDF set. The yellow band corresponds to the ``LHeC 1$^{\mathrm{st}}$ run'' PDFs, while the dark blue shows the final ``LHeC inclusive'' PDFs based on the full LHeC data sets. For the purposes of illustrating the improvement to the uncertainties more clearly, the central values of the LHeC PDFs have been scaled to the CT18 PDF, which itself is displayed by the green band.These plots are taken from Ref.~\cite{LHeC:2020van}.} 
\label{fig:valence}
\end{figure}

Figure~\ref{fig:valence} illustrates the precision that can expected for the valence quark distributions, where it is compared to a selection of modern PDF sets, namely ABMP16~\cite{Alekhin:2017kpj}, NNPDF4.0~\cite{NNPDF:2021njg}, CT18~\cite{Hou:2019efy}, MSHT20~\cite{Bailey:2020ooq}, HERAPDF2.0~\cite{H1:2015ubc} and PDF4LHC21~\cite{PDF4LHCWorkingGroup:2022cjn}.
While the improvement in the determination of the $u_{V}$ distribution is not substantial, a striking advance in $d_{V}$ is visible when adding the LHeC data. In particular, the strong constraints to the very high-$x$ regime are due to the high integrated luminosity.  Note that the yellow band, which displays the ``LHeC 1st run'' PDF, includes only electron and no positron data. Indeed, differences access to valence quarks at low-$x$ can be obtained from the $e^{\pm}p$ cross section. \\
The precise determinations of the $u_{V}$ and $d_{V}$ PDFs in the high-$x$ regime is particularly relevant for Beyond the Standard Model (BSM) searches.A precise determination of the valence quarks distributions will allow the discrimination about current conflicting theoretical pictures for the $d_{V}/u_{V}$ ratio, which cannot be determined precisely with present data, which are inconclusive statistically and suffer from large uncertainties from the use of DIS data on nuclear targets.
\subsubsection{Anti-quarks distributions}
With LHeC data, our knowledge about the sea-quark PDFs will be changed completely. 
\begin{figure}[t!]
\begin{center}
\includegraphics[width=0.95\textwidth]{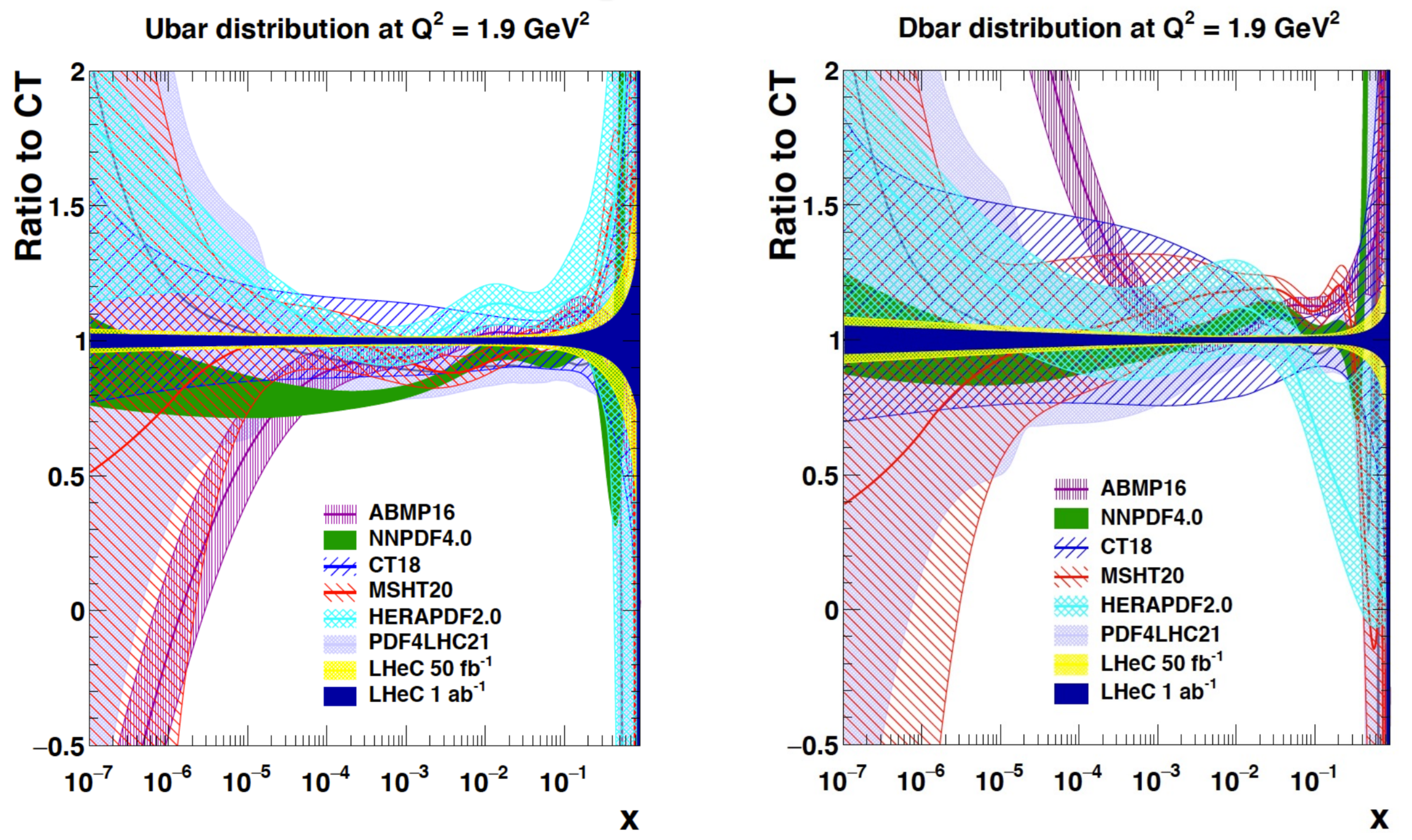}
\end{center}
\caption{Anti-quark distributions at $Q^{2}$ = 1.9~GeV$^{2}$ as a function of $x$, displayed as a ratio to the CT18~\cite{Hou:2019efy} PDF set. The yellow band corresponds to the ``LHeC 1$^{\mathrm{st}}$ run'' PDFs, while the dark blue shows the final ``LHeC inclusive'' PDFs based on the full LHeC data sets. For the purposes of illustrating the improvement to the uncertainties more clearly, the central values of the LHeC PDFs have been scaled to the CT18 PDF, which itself is displayed by the green band.These plots are taken from Ref.~\cite{LHeC:2020van}.} 
\label{fig:sea}
\end{figure}
The $x\bar{U}$ and $x\bar{D}$ PDFs from the 1$^{\mathrm{st}}$ run and the ``LHeC inclusive'' are shown in Figure~\ref{fig:sea} for $Q^{2}$ = 1.9~GeV$^{2}$. A remarkable increase in the determination of the $x\bar{U}$ and $x\bar{D}$ distributions is visible, and it persists from low- to high-$x$. \\
However, the relative uncertainties for $x\geq$ 0.5 are large. This is not particularly worrisome, since the sea-quark contributions are already very tiny in that region. The value of the full LHeC data sample is recognisable in the high-$x$ regime, while the uncertainties of both the small and the full data sets are comparable (and very small) for $x\lesssim$ 0.1.
\subsubsection{The gluon distribution}
\begin{figure}[t!]
\begin{center}
\includegraphics[width=0.95\textwidth]{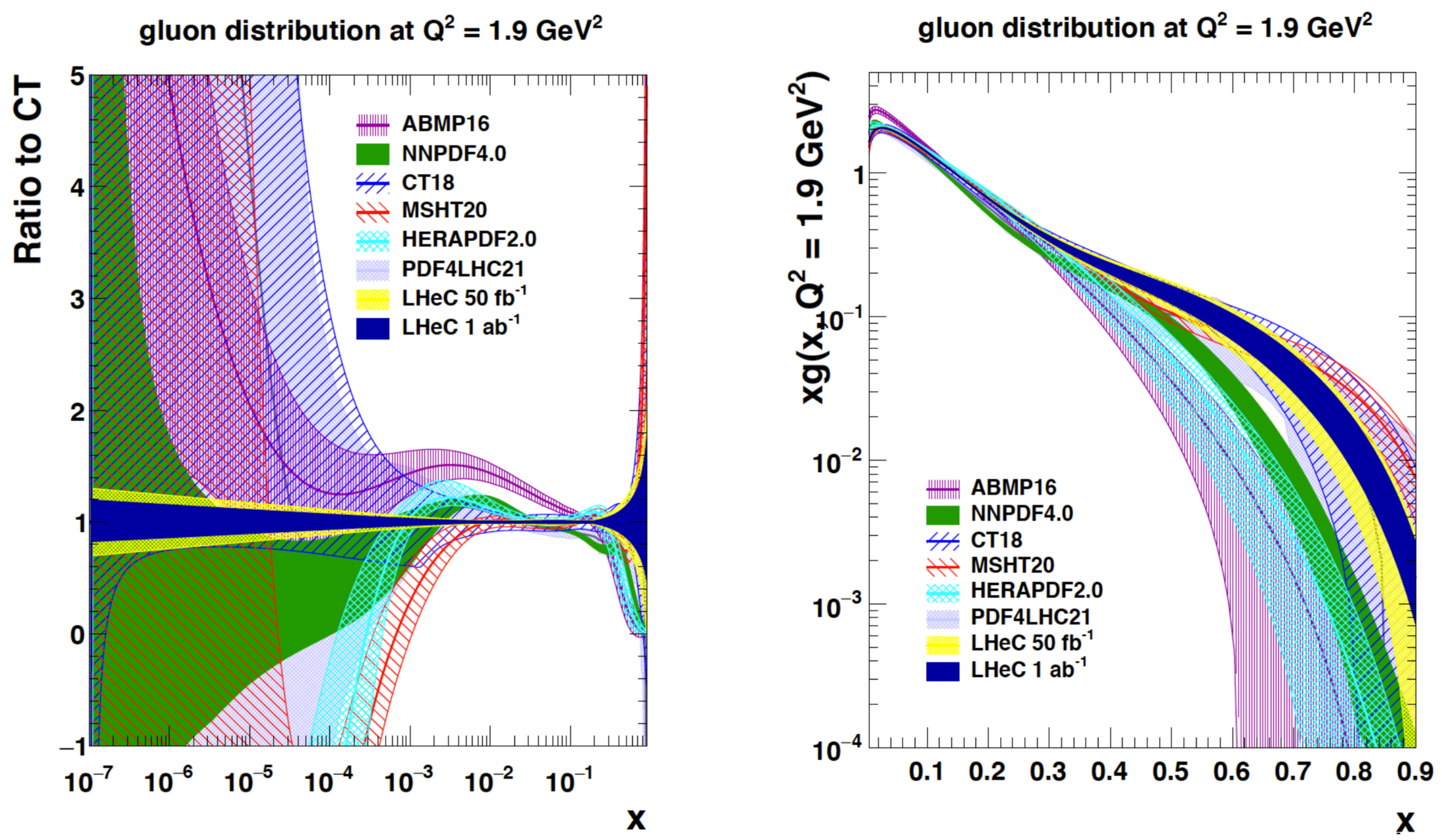}
\end{center}
\caption{Gluon distribution at $Q^{2}$ = 1.9~GeV$^{2}$ as a function of $x$, displayed as a ratio to the CT18~\cite{Hou:2019efy} PDF set. The yellow band corresponds to the ``LHeC 1$^{\mathrm{st}}$ run'' PDFs, while the dark blue shows the final ``LHeC inclusive'' PDFs based on the full LHeC data sets. For the purposes of illustrating the improvement to the uncertainties more clearly, the central values of the LHeC PDFs have been scaled to the CT18 PDF, which itself is displayed by the green band.Left: the distribution is shown on a logarithm $x$ scale and highlights the low-$x$ region. Right: the distribution is shown on a linear $x$ scale and highlights the high-$x$ region. These plots are taken from Ref.~\cite{LHeC:2020van}.} 
\label{fig:gluon}
\end{figure}

With extended kinematic range of DIS and hugely increased precision, the LHeC data can advance our knowledge of the gluon PDF, $xg$, and pin it down much more accurately than it is currently known.\\ Figure~\ref{fig:gluon} shows the gluon distribution,  as it is obtained from the fit to the LHeC inclusive NC and CC data.
When including the LHeC NC and CC precision data, the determination of the gluon PDF will improve dramatically, from the very low-$x$ values,  $\geq$ 10$^{-5}$, to large $x\leq$ 0.8. It is visible how a precision of a few per cent is achieved down to $x\sim$ 10$^{-5}$ can be achieved, and this will help in resolving the question of non-linear parton interactions at small-$x$. \\
Furthermore, in the large-$x$ region,  $x\gtrsim$ 0.3, the very large luminosity provides NC and CC data to accurately access the highest values of $x$, disentangling the sea from the dominant valence part. Thanks to the seminal coverage from very small values of $x$ up to $x\sim$ 1, the gluon PDF can be largely constrained through the momentum sum-rule, and the resulting small uncertainties in the high-$x$ regime are of great importance from BSM searched in $pp$ collisions.
\subsubsection{Parton luminosities}
It is extremely convenient to express the usefulness of the PDF determinations performed at the LHeC with the so-called ``parton luminosities''. These are defined as follows:
\begin{equation}
L_{ab}(M_{X}) = \int{dx_{a}dx_{b}\sum_{q}F_{ab}\delta(M_{X}^{2}-sx_{a}x_{b})}
\end{equation}
where $F_{ab}$ for $(a,b)=(q\bar{q})$ is defined as
\begin{equation}
F_{q\bar{q}} = x_{1}x_{2}\cdot\left[q(x_{1},M^{2})\bar{q}(x_{2},M^{2})+\bar{q}(x_{1},M^{2})q(x_{2},M^{2})\right],
\end{equation}
and $(a,b)$ could also be $(g,q)$, $(g,\bar{q})$ and $(gg)$, without a sum over quarks in the latter case. 
\begin{figure}[t!]
\begin{center}
\includegraphics[width=0.95\textwidth]{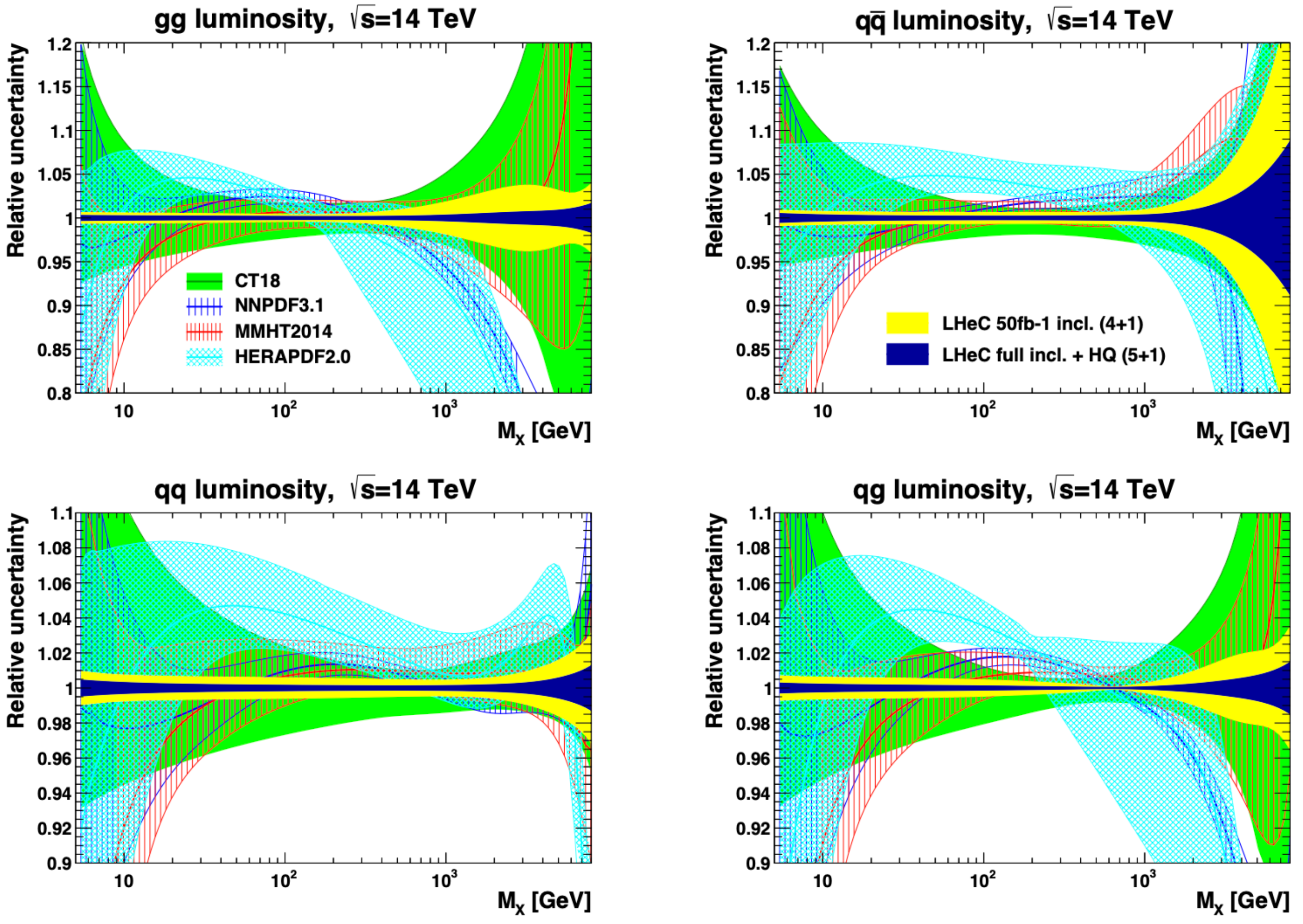}
\end{center}
\caption{Uncertainty bands for parton luminosities as a function of the mass $M_{X} = \sqrt{sx_{1}x_{2}}$ for LHC energies. The yellow band corresponds to the ``LHeC 1$^{\mathrm{st}}$ run'' PDFs, while the dark blue shows a fit to the LHeC inclusive data sets. Both LHeC PDFs shown are scaled to the central value of CT18 PDF set~\cite{Hou:2019efy}. These plots are taken from Ref.~\cite{LHeC:2020van}.} 
\label{fig:partonlumi}
\end{figure}
Figure~\ref{fig:partonlumi} illustrates the expectations for the quarks and gluon parton luminosities.Very precision predictions in a vast range of $M_{X}$ can be provided by the LHeC.  This basically eliminates the main sources of uncertainties due to currently sizeable PDF errors in precision electroweak (EW) measurements performed at the LHC (i.e. the determination of the $W$ boson mass, $m_{W}$, to be within 10$^{-4}$ uncertainty). Moreover, the gluon-gluon luminosity can be determined at 1\%-level for the Higgs mass $M_{X}=M_{H}\backsimeq$ 125~GeV, a factor two or three better than current determinations.

\section{Precise determination of the strong coupling constant}
Since jet cross sections are proportional to $\mathcal{O}(\alpha_{S})$ already at leading-order (LO) QCD, the measurement of these production cross sections at the LHeC will provide sensitivity to the strong coupling constant $\alpha_{S}(m_{Z})$~\cite{H1:2017bml}.  Jets with transverse momentum, $p_{\mathrm{T}}$, between 3 and 500~GeV will be recorded at the LHeC, where the jet energy scale can be calibrated with very high accuracy. It can reach an uncertainty significantly smaller than present LHC experiments, ranging from 0.3\% up to 0.5\%. This can be directly translated into an overall uncertainty of about 5\% (at the very most) on the jet cross section in the Breit frame~\cite{LHeC:2020van}.\\
Performing simultaneous extractions of PDFs+$\alpha_{S}$ when fitting inclusive DIS and jet data together, an uncertainty of $\delta\alpha_{S}(m_{Z})=\pm$ 0.00018~\cite{LHeC:2020van} can be achieved. This values is smaller than the present would average by a more than a factor of five, and it will be a challenge to have equally accurate theoretical predictions at N$^{3}$LO or even beyond.\\
Other related measurements of the hadronic final states, such as multi-jet cross sections, jet substructure observables, $n$-jettiness observables or event shapes, could be included in future PDF fits, and commonly enlarge the sensitivity to the gluon PDF. Moreover, further sensitivity to Transverse Momentum Dependent (TMD) effects can be achieved analysing precision measurements of lepton-jet decorrelation observables~\cite{Liu:2018trl,H1:2021wkz}.

\section{Investigating new small- and high-$x$ dynamics}
Due to the high $ep$ center-of-mass energy, the large acceptance of the LHeC detector and the expected high luminosity of the accelerators, the physics in the low-$x$ regime can be accurately studied at the LHeC. In contract to LHC $pp$ data, since only a single hadron is involved in the collisions happening at the LHeC, the PDF determinations are free from low-$x$-high-$x$ correlations, and the physics phenomena  in these two extreme regions can be studied separately, with high precision.\\
In the large-$x$ region, the higher twists effects become negligible thanks to the ample statistics at such a high $Q^{2}$.  This is particularly relevant for constraining BSM signatures with large mass scales.\\
At small-$x$, the sea quark and the gluon densities rise so much that non-linear and possibly saturation effects may become manifest, as already observed at HERA. These nouvelle dynamics can be studied  in $ep$ and $eA$ collisions at the LHeC for the first time in a reliable way, given that the strong coupling constant is small at such a high scales. This might replace the DGLAP evolution by non-linear evolution and/or BFKL-type equations, with major consequences for physics measurements performed at future hadron colliders (HL-LHC and beyond)~\cite{Bonvini:2018ixe, Bonvini:2018iwt}.  With new measurements of diffractive DIS cross sections, the field of diffractive PDFs will gain new interest~\cite{Armesto:2019gxy}.

\section{Conclusion}
This proceeding illustrates the 
Undoubtedly, our present knowledge of the proton structure has been driven by measurements performed at HERA. With its 1000 times larger luminosity (and higher kinematic reach, as well as center-of-mass energy), the LHeC will equally provide relevant experimental data for precision PDF physics. Furthermore, such independent PDFs are of crucial importance to achieve the physics goal of the HL-LHC programme. Moreover, it has been shown how the large luminosity of the LHeC provides high experimental precision at high $x$, allowing a determination of the strong coupling constant $\alpha_S$ at 0.1\%-level. Furthermore, non-linear and possibly saturation effects can be nicely studied in the low-$x$ region.

\end{document}